\documentclass[a4paper,onecolumn,10pt]{edpsci}
\usepackage{amsmath}
\usepackage{graphicx}
\usepackage{epstopdf}
\usepackage[numbers]{natbib}
\usepackage{hyperref}
\usepackage{url}
\usepackage{comment}
\usepackage{subcaption}
\usepackage{svg}
\usepackage{float}

\hypersetup{colorlinks=true,citecolor=blue,urlcolor=blue,linkcolor=blue}

\usepackage[ruled]{algorithm2e}
\journalname{Journal Title}

\articletype{Research article}

\begin{document}

\title{A multiscale numerical approach to investigate interfacial mass transfer in three phase flow: application to metallurgical bottom-blown ladles}

\titlerunning{Numerical study of mass transfer in metallurgical bottom-blown ladles}


\author{Stefano De Rosa \inst{1,2}$^{,\ast\ast}$ 
\and
Jacob Maarek\inst{2}\correspondingauthor{\email{jacob.maarek@sorbonne-universite.fr}}$^{,}$\thanks{These authors contributed equally to this work}
\and
Stéphane Zaleski\inst{2}}

\authorrunning{S. De Rosa, J. Maarek, S. Zaleski}


\institute{Università degli Studi di Napoli Federico II, Corso Umberto I, 40, 80138 Napoli, Italy \and
Sorbonne Université and CNRS, UMR 7190, Institut Jean Le Rond d’Alembert, 75005 Paris, France}

\abstract{
We use direct numerical simulation (DNS) to investigate mass transfer between liquid steel and slag during a metallurgical secondary refinement process through two reduced-scale water experiments, which reproduce the dynamics seen in an industrial bottom-blown ladle. A container is filled with water and topped by a thin layer of oil, representing the molten steel and slag, respectively. The system is agitated by a bubble plume that impinges on the oil layer and forms an open-eye. A tracer species, dissolved in the water, acts as a passive scalar that is progressively absorbed into the oil layer. 
Both the hydrodynamics and mass transfer in the system are studied and compared with experiments from the literature for ladles of different size and geometry.

The numerical simulation of mass transfer is challenging due to the high Péclet number, leading to extremely thin species boundary layers at the interface. Resolving the boundary layer is  prohibitive even with adaptive grid techniques. A subgrid-scale (SGS) boundary layer model was employed to correct the scalar transport equation, allowing us to solve convection-dominated transport on relatively coarse grids. The hydrodynamics is investigated, and we analyze how the resultant flow field governs mass transport. The numerical results recover two flow regimes: a quasi-steady regime at low flow rates with small deformations of the oil-water interface and an atomising regime at large flow rates. Interfacial species transport is determined to be dominated in an annulus surrounding the open eye caused by a shear layer at the oil-water interface. It is observed that we achieve grid-independent macroscopic quantities that match relatively well with those observed in experiments, allowing use of simulation techniques as a complementary tool going forward.

}





\maketitle

\section{Introduction}
In ladle metallurgy, a reactor, usually of conical or cylindrical shape, is filled with a molten metal bath which is covered by a slag layer. Metallurgical ladles are used in secondary refinement to promote reactions such as desulfurization from the molten metal bath \cite{holappa1982ladle,szekely2012ladle}. Optimal control over these reactions is required to attain the desired steel composition. Reaction kinetics in the ladles are very fast due to high temperatures such that the limiting phenomena for interfacial transport is the diffusion of chemical species across the liquid metal-slag interface. To enhance mixing, a metallurgical ladle is stirred via the injection of an inert gas. A bubble plume impinges on the slag layer, creating a slag-free zone, referred to as the open eye, where the liquid metal reaches the free surface. 

Performing experimental trials in the metallurgical industry is challenging because of high temperatures and the opaque nature of molten metals, consequently requiring specialized measurement techniques.  For these reasons, industrial trials on the actual reactors are very expensive and practically unfeasible. Experimental campaigns are hence generally performed on reduced-scale model ladles and then a similarity criterion is used to grant a relation with the larger industrial ladles. While some studies use liquid metals at lower temperatures \cite{ishida1981effects, hirasawa1987rate, lachmund2003slag} some others further simplify the problem using fluids such as water and oil at room temperature \cite{kim1987physical, de2017mass_article, joubert2021}. In this case, a tracer species that is miscible in oil and water is chosen that mimics the species transfer kinetics observed in the industrial ladle. Thymol is used as the diffusive species in the experiments, as it miscible in both water and the chosen oil, has an equilibrium partition ratio between oil and water similar to that for sulfur between slag and steel, and interfacial mass transfer rate is rate-limited by water-side diffusion, such that the water system mimics the industrial system\cite{kim1987physical}.

Measuring local quantities in the fluid system requires more sophisticated measurement techniques, so the flow has historically been studied by measuring and analyzing a few global quantities: the open-eye size, the fragmentation of the slag layer and the rate of mass transfer. The latter has been thoroughly investigated by Conejo \cite{conejo2020physical} and Liu\cite{liu2024development} by studying the effect of various physical and geometrical parameters on the mass transfer.
Numerical simulations, on the other hand, can greatly improve the understanding of the fluid dynamics in gas-stirred ladles since they allow for easier measurements of local quantities and better visualisation of the flow behaviour. 
While the first simulations on the topic date more than 20 years back \cite{guo2000modeling, alexiadis2004spot}, with the computational power available today increasingly complex simulations have become possible \cite{liu2019review}. 
In particular, while the earliest simulations mainly focused on the flow patterns in the ladle, linking it to the mass transfer by means of models and theories, more recent studies started to address the mass transfer problem itself \cite{senguttuvan2017modeling, joubert2021}. There are however still some gaps in the understanding of the problem.
Furthermore, most of the studies involving numerical simulations do not resolve the turbulent flow explicitly, but they rather use turbulence models such as large eddy simulations or Reynolds-averaged Navier-Stokes to treat smaller scale phenomena. Only recently some studies on ladles have been performed using direct numerical simulations (DNS) that solve turbulence to the smallest scales \cite{li2021toward, li2023large}. Lastly, the majority of numerical studies has focused on the hydrodynamic problem and ignores resolving mass transfer. 

The first numerical study to simulate both the flow and mass transfer was the work of De Olivera Campos et al. \cite{de2017mass_article}, quickly followed by the works of Joubert et. al. \cite{jothese, joubert2021}. Both works also performed analogous experiments to directly compare with the performed simulations. In these works, mass transfer is simulated by artificially increasing the diffusivity such that the Schmidt number $Sc = \mu/(\rho D)$ is of order one. To compare with experiments, the results are extrapolated to the physical $Sc$ number using correlations from mass transfer theory.
The applicability of this approach is questionable as the simulations are performed  in a regime where mass and momentum boundary layers are of comparable magnitude, whereas the real system is governed by convection-dominated transport with very thin species boundary layers. 
Consequently, a large uncertainty is reported for the simulated mass transfer rates. This work builds on the previous numerical work of Joubert et al, with the principal improvement being the addition of a multiscale mass transfer model to simulate mass transfer without artificially increasing diffusivity.  
We chose to reproduce two experimental campaigns: the one by Joubert et al.  \cite{joubert2021} and the one performed by Kim \& Fruehan \cite{kim1987physical}.  We simulate the system and investigate the relationship between the open-eye size and mass transfer rate for an input gas flow rate.  The numerical simulations were performed using the free software platform Basilisk, a PDE solver using adaptive Cartesian grids extensively employed for modelling multiphase flows \cite{basilisk, popinet2003}. The results are demonstrated to be grid-convergent for the predicted mass transfer rate in relative agreement with the published experimental results. Importantly, the multiscale mass transfer model makes resolving the scalar-transport problem of comparable computational cost with respect to the cost of the flow simulations.

\section{Material and Methods}

\subsection{Model description}
\begin{figure}[t]
    \centering
    \includegraphics[width=0.7\linewidth]{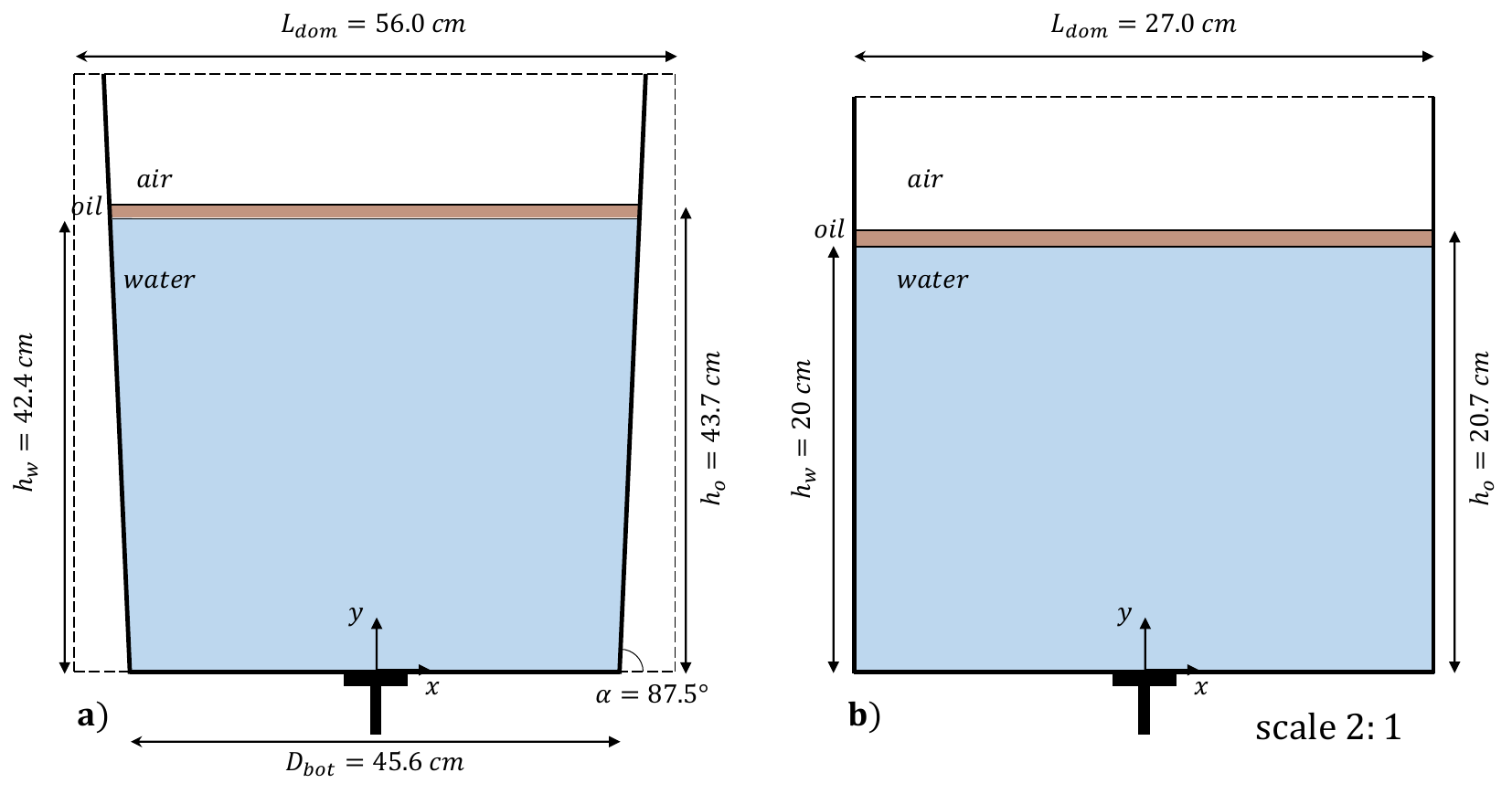}
    \caption{Geometry and initial conditions for the two ladles studied: a) the conical ladle of Kim \& Fruehan \cite{kim1987physical}; b) the cubic ladle of Joubert et al. \cite{joubert2021} (in scale 2:1 with respect to panel a). }
    \label{fig:ladle_scheme}
\end{figure}

The two ladles that we analysed are shown in Fig.~\ref{fig:ladle_scheme}: in panel a) the truncated conical ladle studied by Kim \& Fruehan \cite{kim1987physical} and in panel b) the cubic ladle of the experiments by Joubert \cite{joubert2021}, presented at 2:1 scale with respect to panel a. In the former, the dotted lines represent the computational domain that encloses the ladle since the software tool is limited to cubic domains. 

At initial time, the three fluids are still and stacked on top of one another: water, then the oil mixture and then air. Air is injected from the bottom of the ladle with varying flow rates in the different simulations.  Bubbles form and start to rise in the water until they reach the water-oil and oil-air interfaces, forming the open eye and merging into the air above. 
The outflow boundary is far enough from the oil surface to ensure that only air can flow out of the domain, guaranteeing the total volume of the fluids to be constant throughout the simulations. 

The simulations are performed in two steps: in the first only the hydrodynamics is active while in the second step mass transfer is turned on. To remove a transient state, the hydrodynamic simulations are run until a pseudo-steady state condition for measured quantities, e.g. kinetic energy, droplet size distributions and open-eye size, is attained.  Subsequently, a passive scalar tracer is initialized uniformly in the water phase, while its value is set to zero in oil. The mass transfer simulations are performed until an equilibrium state is observed for the global mass transfer rate.

To relate the flow rates between industrial and reduced scale ladles, we have used a similitude criterion based on a modified Froude number commonly used in the literature \cite{liu2019review}, 

\begin{equation}
N = \left( \frac{Q^2}{g h_{\mathrm{w}}^5} \right)^{1/10}, \label{eq:Froude}
\end{equation} 
where $Q$ is the flow rate, $g$ is the acceleration due to gravity, and $h_{\mathrm{w}}$ is the initial height of the water layer. It is reported that the change in the open-eye area is approximately linear with respect to $N$.

The relevant dimensionless number to characterize the mass transfer between water and oil is the global Sherwood number. It compares the rate of the total mass flux to the diffusive mass flux and is defined as
\begin{equation}
    \overline{Sh} = \frac{K_{\mathrm{w}} A h_{\mathrm{w}}}{L_{\mathrm{x}}^2 D_{\mathrm{w}}},
    \label{eq:Sherwood}
\end{equation}
where $K_{\mathrm{w}}$ is the mass transfer coefficient, $A$ is the exchange area between water and oil, $D_{\mathrm{w}}$ is the diffusivity in the water and $L_{\mathrm{x}}$ is the characteristic dimension which we take to be the diameter of the bottom of the oil layer.

All numerical values used in the simulations are found in Table~\ref{tab-1}.
It is worth mentioning, as it may lead to some discrepancies between the experimental and numerical results that while for the ladle of Joubert's experiments all dimensions and numerical values were available in the paper, for the ladle employed by Kim \& Fruehan some values were missing or uncertain.
In particular, there was a discrepancy between the volumes of water and oil indicated in their paper and the fluids heights provided that, together with the dimensions of the ladle, would lead to slightly different volumes. We have chosen to use the fluid heights shown in Tab.~\ref{tab-1} to ensure volumes of water and oil of $V_{\mathrm{w}}=75 \ $L and $V_{\mathrm{o}}=2.5 \ $L. Furthermore the surface tension between oil and air was not present in the paper, so we choose to use the same value as in Joubert's experiments. 
Of all the process conditions investigated by Kim \& Fruehan in their experiments, we analysed the one they have used as a reference: the "light paraffin oil $+$ cottonseed oil" mixture with air injection by means of a single centred tuyere of diameter $4.8$ mm. 

\begin{table*}
\caption{Numerical values of the geometrical parameters and physical properties for the two ladles in analysis.}\label{tab-1}
\centering
\begin{tabular}{|cccc|cccc|}
\hline
Variable  & Units  & Kim \& Fruehan & Joubert et al. &  Variable  & Units  & Kim \& Fruehan & Joubert et at\\
\hline
$h_{\mathrm{w}}$ & cm & 42.4 & 20.0      & $h_{\mathrm{o}}$ & cm & 43.7 & 20.7 \\
$L_{\mathrm{d}}$ & cm & 56.0 & 27.0    & $d_{\mathrm{in}}$ & mm & 4.76 & 7.90 \\
$R_{\mathrm{b}}$ & cm & 27.8 & /       & $\alpha$ &  & 87.5° & / \\
$\rho_{\mathrm{a}}$ & kg/m$^3$ & 12.25 & 12.25 & $\mu_{\mathrm{a}}$ & Pa $\cdot$ s & 1.85e$^{-5}$ & 1.85e$^{-5}$ \\
$\rho_{\mathrm{w}}$ & kg/m$^3$ & 998 & 998 & $\mu_{\mathrm{w}}$ & Pa $\cdot$ s & 1.002e$^{-3}$ & 1.002e$^{-3}$ \\
$\rho_{\mathrm{o}}$ & kg/m$^3$ & 886 & 920 & $\mu_{\mathrm{o}}$ & Pa $\cdot$ s & 3.34e$^{-2}$ & 7.90e$^{-2}$ \\
$\sigma_{\mathrm{aw}}$ & N/m & 7.2e$^{-2}$ & 7.2e$^{-2}$ & $\sigma_{\mathrm{ao}}$\tablefootmark{1} & N/m &  3.17e$^{-2}$ & 3.17e$^{-2}$ \\
$\sigma_{\mathrm{wo}}$ & N/m & 1.81e$^{-2}$ & 2.55e$^{-2}$ & P  & & 350 & 350  \\
$D_w$ & m$^2$/s & 6.8e$^{-10}$ &6.8e$^{-10}$   & $D_o$\tablefootmark{1} & m$^2$/s & 6.8e$^{-12}$ & 6.8e$^{-12}$ \\

\hline
\end{tabular}
\tablefoot{ \\
\tablefoottext{1}{In the Kim \& Fruehan paper the values for $\sigma_{\mathrm{ao}}$ and $D_o$ are not reported, so we have used the same value for both cases.}
}
\end{table*}

\subsection{Mathematical model}
\label{sec-math-model}
The experiments analyzed are isothermal and they involve three immiscible phases: water, air and oil. While water and oil are incompressible, the injected air bubbles are in theory compressible. To simplify the model and reduce the computational load, we neglect the compressibility effects of air. This hypothesis is reasonable if the hydrostatic pressure of the system is significantly smaller than the atmospheric pressure $ \rho_{\mathrm{w}} g h_{\mathrm{w}} \ll p_{\mathrm{atm}} $. 

Esmaeeli \& Tryggvason \cite{esmaeeli1998direct} and Cano-Lozano et al. \cite{cano2016paths} studied the effect of the numerical value of the air density in bubbly flows. They reported that a higher numerical air density had only a small influence on the rise velocity, which is explained by the pressure gradient in the bubble remaining much smaller than in the surrounding liquid. We also show in Appendix \ref{bubble_plume_structure} that momentum-driven effects should have negligible effect on the vertical velocity of the bubble plume. Consequently, we can increase the air density with negligible effect on the flow.  A reduction of the density ratio $\rho_{\mathrm{w}}/\rho_{\mathrm{a}}$, makes the pressure-Poisson problem converge significantly faster, increasing computational efficiency. We choose to use a value of the air density $\rho_{\mathrm{a}}$ that is ten times bigger than the air density at STP, as shown in Table \ref{tab-1}.

Interface capturing is included with the multi-VOF approach, with three indicator functions $f_i$ to identify the different phases. The $f_i$ have to respect a volume conservation equation
\begin{equation}
\frac{\partial f_i}{\partial t} + \nabla \cdot ( \mathbf{u} f_i ) = 0. \label{Hadv}
\end{equation}
The three $f_i$ have to sum up to 1 in each cell at each time instant
\begin{equation}
 \sum_{i=1}^3 f_i = 1. 
\end{equation}
Local physical properties are computed with the VOF-weighted average of the phasic properties.

Under the formulated hypotheses, we write the incompressible Navier-Stokes equations for the three-phase problem using the one-fluid method as
\begin{equation}
\nabla \cdot \mathbf{u} = 0, \label{divu}
\end{equation}
and
\begin{equation}
\rho \left( \frac{\partial \mathbf{u}}{\partial t}  +  \mathbf{u} \cdot \mathbf{\nabla} \mathbf{u} \right) = - \mathbf{\nabla} p + \mathbf{\nabla} \cdot \tau + \mathbf{f_{\mathrm{c}}} + \rho \mathbf{g}, 
\label{nseq_0}
\end{equation}
where $\rho(\mathbf{x},t)$ is the fluid density, $\mathbf{u}$ the velocity, $p$ the pressure, $\tau $ the viscous stress tensor for incompressible flow given by
\begin{equation}
\tau  = \mu \left( \mathbf{\nabla} \mathbf{u} + (\mathbf{\nabla} \mathbf{u} )^T \right), \label{tauvdef}
\end{equation}
$\mathbf{g} = -g \mathbf{e}_y$ is the acceleration of gravity, and $\mathbf{f}_{\mathrm{c}}$ is the capillary force. This force depends on the local interface: the domain contains three phases that are separated by sharp interfaces $S_{ij}$, with the indexes $i,j\in[1,3]$ corresponding to the three fluids. The capillary force is computed with the continuum surface tension force (CSF) method, which for a three-phase system is computed as 
\begin{equation}
    \mathbf{f}_{\mathrm{c}} = \sum_{i=1}^3 \sigma_i \kappa_i \nabla f_i,
\end{equation}
where $\sigma_i$ are phasic surface tensions defined as $\sigma_i= 0.5 (\sigma_{ij}+\sigma_{ik}-\sigma_{jk})$, while $\kappa_i$ is the interface curvature of the $i^{th}$ phase and $f_i$ its volume fraction. The phasic curvature is computed with a hybrid approach developed in \cite{zhao2024hybrid}, where away from the three--fluid contact line the height function method \cite{POPINET20095838} is used and at the contact line the phasic curvature is computed by the divergence of the interface normal \cite{brackbill1992continuum}. This approach was optimal for accurate calculation of the net force on the contact line which is necessary for accurate computations in capillary three-phase flows while minimizing parasitic currents. 

The spatial and temporal schemes used for the incompressible Navier-Stokes equations are described in \cite{basilisk, popinet2003, POPINET20095838}. It uses an operator split approach for the momentum equation, with a projection method used to enforce the continuity equation. The advection of each VOF scalar field is performed using the dimension-split geometric scheme of \cite{WEYMOUTH20102853}, which conserves mass to machine tolerance for a divergence-free velocity field.

After phasic advection, the volume fractions are normalized to guarantee that they always add to one. Although this procedure does not conserve mass, it is observed that mass conservation converges at first order with grid refinement. We measured it and confirmed that the ratio of the change in the volume of the fluid phases to the initial phasic volume is of the order of $10^{-3}$ to $10^{-4}$ over the performed time interval for all simulations. 

Species transfer of a tracer between the water and oil phases is described by the scalar transport equation with the diffusive flux calculated by Fick's law

\begin{equation}
\frac{\partial C_i}{\partial t} + \mathbf{u} \cdot \nabla C_i = \nabla \cdot (D_i \nabla C),
\end{equation}
where $C_i$ is the mass concentration in phase $i$ and $D_i$ the diffusivity of the chemical species when dissolved in that phase. There is no species flux at either the air-water or air-oil interfaces, nor in the solid. We consider dilute species transfer, ignoring local phasic volume changes due to species transfer. This assumption leads to a diffusive flux continuity equation
\begin{equation}
D_{\mathrm{o}} \nabla C_{\mathrm{o}} = D_{\mathrm{w}} \nabla C_{\mathrm{w}}.
\label{eq:fluxcont}
\end{equation}
Furthermore, Henry's law holds, stating that the ratio between the oil and water-side concentration is given by a partition ratio $P$

\begin{equation}
C_{\mathrm{o}} = P C_{\mathrm{w}} \label{eq:henryslaw}.
\end{equation}

Numerically, the scalar transport equation is solved using a two-field approach, where we transport the conserved quantities 
\begin{equation}
 g_i = C_i f_i, \label{cons_quant}
\end{equation}
corresponding to the species mass in the oil and water phases. Two-scalar transport equations are solved, one corresponding to mass transfer in the water phase and the other in the oil phase. The phasic scalar transport equation for $g$ is written as 
\begin{equation}
\frac{\partial g_i}{\partial t} + \nabla \cdot (\mathbf{u} \, g_i) = \nabla \cdot (D_i \, A_i \nabla C_i), \label{scalar trans}
\end{equation}
where $A_i$ is the phasic area. The advection term is treated with a geometric scheme consistent with VOF advection and prevents numerical diffusion across the interface, as shown in reference \cite{LOPEZHERRERA201514}. The phasic diffusion term is treated with a time-implicit scheme, and the diffusive flux across the interface is included as equal and opposite explicit source terms in the two phasic diffusion solvers. Species mass is conserved up to the tolerance of the pressure-Poisson solver which enforces continuity.

\subsection{Simulations setup}
All simulations were performed using the open-source platform Basilisk \cite{basilisk, popinet2003, POPINET20095838}, which employs the Finite Volume method to solve transport equations on adaptive Cartesian grids. 
The solver has been thoroughly validated for complex interfacial flows \cite{farsoiya2017axisymmetric, gumulya2021dynamics, mostert2020inertial,berny2020role} and the used code is generally consistent with that used by Joubert \cite{joubert2021, jothese} with modifications to improve the model and include arbitrary geometries.

It was observed that in the simulations by Joubert, the mass and momentum injection were dependent on mesh refinement, with an under prediction of the effective air flow rate ranging between 1 and 10$\%$ depending on the level of refinement.  This error had two causes: a second order approximation for the geometry of the injector and a miscalculation in the treatment of phasic VOF-advection from the ghost cells at domain boundaries. We solved the errors, resulting in a mesh-independent mass and momentum flux, using  a small modification of VOF and mass-momentum consistent advection algorithms \cite{vaudor2017consistent,arrufat2021mass} and the VOFI library \cite{bna2016vofi} to correct the injector geometry.

Furthermore, the Joubert simulations underestimated the spreading force at the three-phase contact line, which caused the open-eye area to be over-predicted by up to 30 percent compared to experiments. The improved surface tension calculation due to the hybrid curvature calculation described in \ref{sec-math-model} led to the formation of a thin oil meniscus that significantly reduced the size of the open eye, physically consistent with the positive spreading coefficient $S_o = \sigma_{aw} - \sigma_{ao} -\sigma_{ow}$ of the oil-water-air system. However, meniscus coverage was observed to be grid dependent, as coarse grids could not capture the thin film. Furthermore, comparisons to experiments would be difficult in the presence of thin film, as sophisticated optical tools were not used when measuring the open-eye in experiments. Using the meniscus area coverage as an error bar on the open-eye area in simulations led to relative agreement with experiments (see \cite{maarekthese}).

Finally, the numerical experiments of Joubert used an immersed Dirichlet boundary condition at the oil-water interface when computing mass transfer, effectively considering the oil phase as an infinite well. We perform a coupled mass transfer simulation with equilibrium transmission boundary conditions at the interface (see equations \ref{eq:fluxcont} and \ref{eq:henryslaw}), which allows us to consider effects such as the saturation of the species in oil.

\subsubsection{Implementing arbitrary solid geometries}
In order to describe the truncated cone ladle on a structured Cartesian grid, we use the fictitious solid method \cite{tryggvason2011direct}. This approach consists in having the fluid phases fill the whole cubic domain and adding a source term to the momentum equation such that the velocity field is null inside the solid. In practice, this is accomplished using a mask and overwriting the velocity field with
\begin{equation}
    \mathbf{u}(\mathbf{x})=I(\mathbf{x})\mathbf{u}_{\mathrm{f}}(\mathbf{u}) + \left( 1-I(\mathbf{x}) \right) \mathbf{u}_{\mathrm{s}}(\mathbf{u_x}),
\end{equation}
where $I(\mathbf{x})$ is an indicator function that is equal to 1 inside the fluids and 0 inside the solid. Since the ladle is static, the solid velocity $\mathbf{u}_{\mathrm{s}}$ is null and the equation reduces to

\begin{equation}
    \mathbf{u}(\mathbf{x})=I(\mathbf{x})\mathbf{u}_{\mathrm{f}}(\mathbf{u}),
\end{equation}
The described approach is equivalent to using Brinkman penalization with the penalization coefficient equal to zero \cite{https://doi.org/10.1002/fld.307}. 

Numerical slip of the fluid-fluid-solid contact line was observed to be required to suppress spurious currents generated from the chosen solid model. This was accomplished by a regularization of the fluid interface at the contact line. First, we extrapolate the fluid interface horizontally into the solid by setting the fluid volume fractions in the two neighbouring solid cells to match the values of the closest cell containing a fluid-solid interface. For a vertical cylinder, an approximation of the truncated conic geometry, this extrapolation would impose 90° contact angles. Second, we avoid the pinning of the contact line by checking if the neighbouring cells on the interior of the solid interface are part of the bulk of a fluid and, if that is the case, we overwrite the volume fraction in the fluid-solid mixed cell to be filled by that same fluid. This method was easy to implement and only leads to a negligible loss of mass conservation at the solid boundary contact line. 

This section is not relevant for the simulations of the Joubert case as the cubic ladle matches the native cubic domain used in Basilisk. 

\subsubsection{Boundary conditions}

 Inflow and outflow boundary conditions are used at the bottom and top of the domain and no-slip boundary conditions are imposed on the sides. A uniform inlet velocity is applied at the injector and is treated with an indicator function $I_{\mathrm{inj}}$ with the velocity given by: 
 \begin{equation}
     U_{\mathrm{b}} = \frac{Q}{A_{\mathrm{inj}}}I_{\mathrm{inj}},
 \end{equation}

 Homogeneous Neumann boundary conditions are applied for the passive scalar on all fluid domain boundaries, which prevents diffusive flux out of the boundary.

\subsubsection{Mesh}
 The domain is spatially discretized using an octree grid adapted in time using the velocity, volume fraction, and concentration fields as adaptation criterion \cite{popinet2015quadtree,van2018towards}. The smallest scale resolved is given by
 \begin{equation}
    \Delta x_{\mathrm{min}}= \frac{L}{2^{\mathrm{M}}},
\end{equation}
where $L$ is the domain size and $\mathrm{M}$ is the maximum level in the hierarchical grid.

A mesh independence study for measured hydrodynamic and mass transfer quantities was performed in \cite{joubert2021, maarekthese} in which the resolution was varied from roughly 500 $\mathrm{\mu m}$ to 125 $\mathrm{\mu m}$. In this paper, we will only present a brief mesh convergence study for the Kim \& Fruehan case for two flow rates. All other simulations on the ladles have been performed using max resolution of roughly 500 $\mathrm{\mu m}$, corresponding to a maximum level of refinement of 10 for the Kim \& Fruehan case and of 9 for the Joubert case.

For the used resolution, the cell count varied from 8 million to 40 million for the Joubert case and 25 million to 125 million for Kim \& Fruehan case depending on the flow rate. Massively parallel simulations were performed with a number of processors ranging from 512 to 8192, with the total cost ranging from roughly 100 thousand CPU hours for the smallest to 3 million CPU hours for the largest simulations.

\subsubsection{Passive Scalar Subgrid-Scale modelling}

In coupled hydrodynamic-mass transfer problems at high Schmidt numbers, the minimum length scale in the problem is the concentration boundary layer at the fluid-fluid interface. This is explained by relating the size of the concentration boundary layer to that of the momentum boundary layer with the relation $\delta_c \sim \delta_u/ Sc^{b}$, where $\delta_c$ is the chemical boundary layer, $\delta_u$ is the momentum boundary layer, $Sc$ is the Schmidt number, and $b$ is a scaling coefficient that varies between 1/3 and 1/2 depending on if the local flow in the reference frame of the moving interface is closer to that of a flow adjacent to a rigid or a free surface \cite{maarekthese}. Using this scaling argument, one could say that for a Schmidt number in the water $Sc = 1480$, used in the simulated system, the species boundary layer would be between roughly 11.4 and 38.5 times smaller than the momentum boundary layer. Furthermore, we estimate the mesh size required to simulate the hydrodynamic and mass transfer system with the estimations $\Delta x_{u} = C_1 \delta_u$ and $\Delta x_{c} = C_2 \delta_c$, respectively, where $C_1$ and $C_2$ are coefficients smaller than one. The actual values of the coefficients are related to the used numerical discretization and are representative of how many cells are required in the boundary layer to achieve grid converged results. It is observed in the used numerical schemes that $C_2 < C_1$, thus further reducing required mesh size for mass transfer simulations. Lastly we can say that in the case of the performed simulations, the computational cost required scales roughly like $T_{\mathrm{sim}} \propto (1/\Delta x)^3 $, where $T_{\mathrm{sim}}$ is the cost of the simulation measured in CPU hours per unit simulated second. Therefore using the estimate $\Delta x_c \approx 45 \Delta x_u$, the cost of the mass transfer simulation would be 100 thousand times the cost hydrodynamic simulation.  The proposed scaling for the computational cost is observed numerically and is due to refinement at the interface increasing the number of cells $N \propto (1/\Delta x)^2$ and the timestep for the considered high Weber number flow and chosen grid size controlled by the CFL condition. 

Fortunately, the scale separation between the mass transfer and hydrodynamic length scales when the $Sc$ number is large allows us to simplify the scalar transport equation near the interface, leading to simple profiles for the species mass distribution near the interface. These profiles derived from boundary layer theory are used to perform subgrid corrections for the convective and diffusive fluxes. An obvious analogue can be made to the use of wall functions in single phase flow to increase the shear stress in the vicinity of the wall. 

To derive a subgrid model, we first assume that the mass transfer resistance in the water layer is much larger than that of the oil layer, known from the global kinetics of the system \cite{kim1987physical}. This allows us to compute the interfacial flux with a one-sided subgrid model in the water phase. 
We consider an advection-diffusion problem entirely located inside the layer $|z - z_{ow}| \ll \delta_{u,w}$, solved in the reference frame of the moving interface. In this sublayer, the velocity field $U$  may be linearized so that $U_r(z,t) = U_\Sigma(t) + \omega(t) (z -  z_{ow} )$, representing an interface slip velocity and a shear dependent term. We ignore tangential diffusion. The described boundary layer approximation leads to the simplified advection-diffusion equation for the concentration field $C(r,z,t)$
\begin{equation}
[ U_\Sigma + \omega (z -  z_{ow} )] \partial_r C = D_w \partial^2_{zz} C, \label{ediff}    
\end{equation}
with Dirichlet boundary conditions at the interface $C(r,0,t)=C_\Sigma(t)$ and in the far field $C(r, \infty, t) =  C_{\infty}(t)$. The concentration at the interface is estimated by performing a one-way extrapolation for the concentration field at the interface on the oil side and applying Henry's law to calculate the concentration on the water side. The far field concentration is estimated as either the average water-phase concentration or initial (uniform) water-phase concentration with negligible effect on results. The generalized Eq. \ref{ediff} was solved by Apelblat \cite{apelblat1980mass} using of a change of variables, giving 
a concentration profile and interfacial mass flux as a function of the nondimensional distance  $R = D_w \, r \,  \omega^2/u_\Sigma^3$. 

For the purposes of simplifying the subgrid model, we will limit ourselves to the limiting cases where either $\omega$ or $u_\Sigma$ are zero, corresponding to a flow tangential to a free or rigid surface, respectively. These simplified problems are equivalent to an unsteady diffusion problem where the time variable is replaced with $r$ and the known L{\'e}v{\^e}que problem \cite{leveque1928lois}, respectively. In these regimes, the solution for Eq. \ref{ediff} gives the concentration profiles
\begin{equation}
    C(r,z) = c_\Sigma + (c_\infty-c_\Sigma)\mathrm{erf}\left( \frac{z \; u_\Sigma^{1/2}}{2 \; D^{1/2} \; r^{1/2}} \right), \label{free_surface_erf}
\end{equation}
and 
\begin{equation}
    C(r,z) = c_\Sigma + (c_\infty-c_\Sigma)\frac{3}{\Gamma(1/3)}\int_0^{z(\omega/(9\, D \, r))^{1/3}}\exp(-t^3)dt. \label{sol_leveque}
\end{equation}
The concentration profiles are written as scale invariant shape functions, where the shape function for the error function solution is written as 
\begin{equation}
    C(\delta_c,z) = c_\Sigma + (c_\infty-c_\Sigma)\mathrm{erf}\left( \frac{z }{\delta_c} \right),
\end{equation}
where $\delta_c$ is the local boundary layer thickness. Integrating the concentration profile over a control volume gives the species mass in the cell, represented by $g$ in Eq. \ref{cons_quant}. Consequently $\delta_c$ and the species mass distribution near the interface are implicitly described by the value $g$ in the interfacial cell, the applied shape function, and the concentration boundary conditions at the interface and in the far field. Knowing the species concentration distribution allows us correct the convective and diffusive fluxes in the transport equation. In the simulation of the Joubert experiment, the L{\'e}v{\^e}que shape function was used, from a hypothesis that the large viscosity ratio between the oil and water would lead to a velocity boundary layer in the water closer to that of a rigid surface. An a posteriori analysis showed that in reality the boundary layer would be expected to be in an intermediate regime but closer to a free surface than a rigid surface \cite{maarekthese}. As such the free surface error function was applied in the Kim \& Fruehan case, where the viscosity ratio is half that in the Joubert case. It was shown in \cite{maarekthese} that the choice of the shape function introduces a maximum error of approximately 12 percent for the SGS flux. 

The described SGS model has been independently developed in the works \cite{ABOULHASANZADEH2013165,weiner_claassen_transient, WEINER2017261, maarekthese}, with principal differences relating to the data structures used to track the interface and represent the species in the chemical boundary layer, and the treatment of the convective term. In the method proposed by \cite{ABOULHASANZADEH2013165} and improved in \cite{weiner_claassen_transient}, the interface is tracked with a Front-Tracking method and the species in the boundary layer is separated from the bulk and stored on the front markers. Secondly the convective term is split in two terms, one treated by the Lagrangian motion of the markers and the second is proportional to the surface divergence of the flow, leading to the compression and expansion of the boundary layer. The second approach, first proposed in \cite{weiner_claassen_transient} and improved in \cite{maarekthese}, applies the SGS model in the geometric VOF framework. This approach is naturally used in the present simulations. Combining the chemical boundary layer shape function with a piecewise linear reconstruction (PLIC) of the interface, we write the species distribution of mass in the boundary layer as
\begin{equation}
\begin{array}{c}
    c^{\mathrm{SGS}}(\mathbf{x}, \mathbf{n}, \alpha, \delta_c)  = 
\left\{
    \begin{array}{lr}
         c_\Sigma + (c_\infty - c_\Sigma ) \mathrm{erf} \left ( \frac{ \mathbf{n} \cdot \mathbf{x} - \alpha}{\delta_{c}} \right) & \text{if } \mathbf{x} \in \Omega_c(t) \\
        0  & \text{otherwise}
    \end{array}
\right\}.
\end{array}
\label{mass_disto_erf}
\end{equation}
The convective flux is computed geometrically consistent with VOF method
\begin{equation}
    F^{\mathrm{SGS}}_{\mathrm{Adv}} = \int_{V_u}  \, c^{\mathrm{SGS}}(\mathbf{x}, \mathbf{n}, \alpha, \delta_c) \, dV, \label{sgs_conv_analytic}
\end{equation}
where relative to VOF phasic advection, the integral of a Heaviside function is replaced with the applied shape function. The diffusive flux is computed by evaluating the surface integral of the concentration gradient
\begin{equation}
    \begin{split}
        F^{\mathrm{SGS}}_{\mathrm{Diff}, x} = -D \, A \; c_x^{\mathrm{SGS}} = & -D \, \int_{A} \partial_x c^{SGS}(\mathbf{x}, \mathbf{n}, \alpha, \delta_c) dA \\&
         \approx -D  \, n_x  \int_{A} \partial_n c^{SGS}(\mathbf{x}, \mathbf{n}, \alpha, \delta_c) dA, \\&
    \end{split}
\label{diffusive_flux_sgs}
\end{equation}
where $\partial_n c^{SGS}$ is the 1-D gradient of the concentration profile with respect to the distance from the interface. To integrate the SGS diffusive flux in a time-implicit diffusion solver, an effective diffusivity is calculated considering the ratio of the SGS flux and the grid-resolved flux
\begin{equation}
    D_\mathrm{eff, x} = D_w \, \frac{c_x^\mathrm{SGS}}{c_x^\mathrm{FD}},
\end{equation}
where $c_x^\mathrm{FD}$ is the partial derivative estimation given by a finite difference.

The technical details and validation of the used algorithm are out of the scope of this work but are found in their entirety in \cite{maarekthese}.

\section{Results}

This section presents the simulation results for both hydrodynamics and mass transfer. Detailed results are provided for the Kim \& Fruehan case, while only principal results are highlighted for the Joubert case, for which additional plots and figures are available in the supplementary materials.

\subsection{Mesh convergence study}
To analyse the effect of mesh refinement on the mass transfer results, we perform a mesh convergence study on the Kim \& Fruehan case at two flow rates of $0.5$ and $2$ L/min. The maximum level of refinement was increased from 10 to 11 in a 5x5x5 stencil centred around the oil-water interface, whereas the mesh was kept the same away from the interface. This was done to save computational resources, but is still expected to capture any change in interfacial mass transfer.  A more complete mesh convergence study for the Joubert case is found in the references  \cite{jothese, joubert2021, maarekthese}.

\begin{figure}[H]
    \centering
    \includegraphics[width=0.65\linewidth]{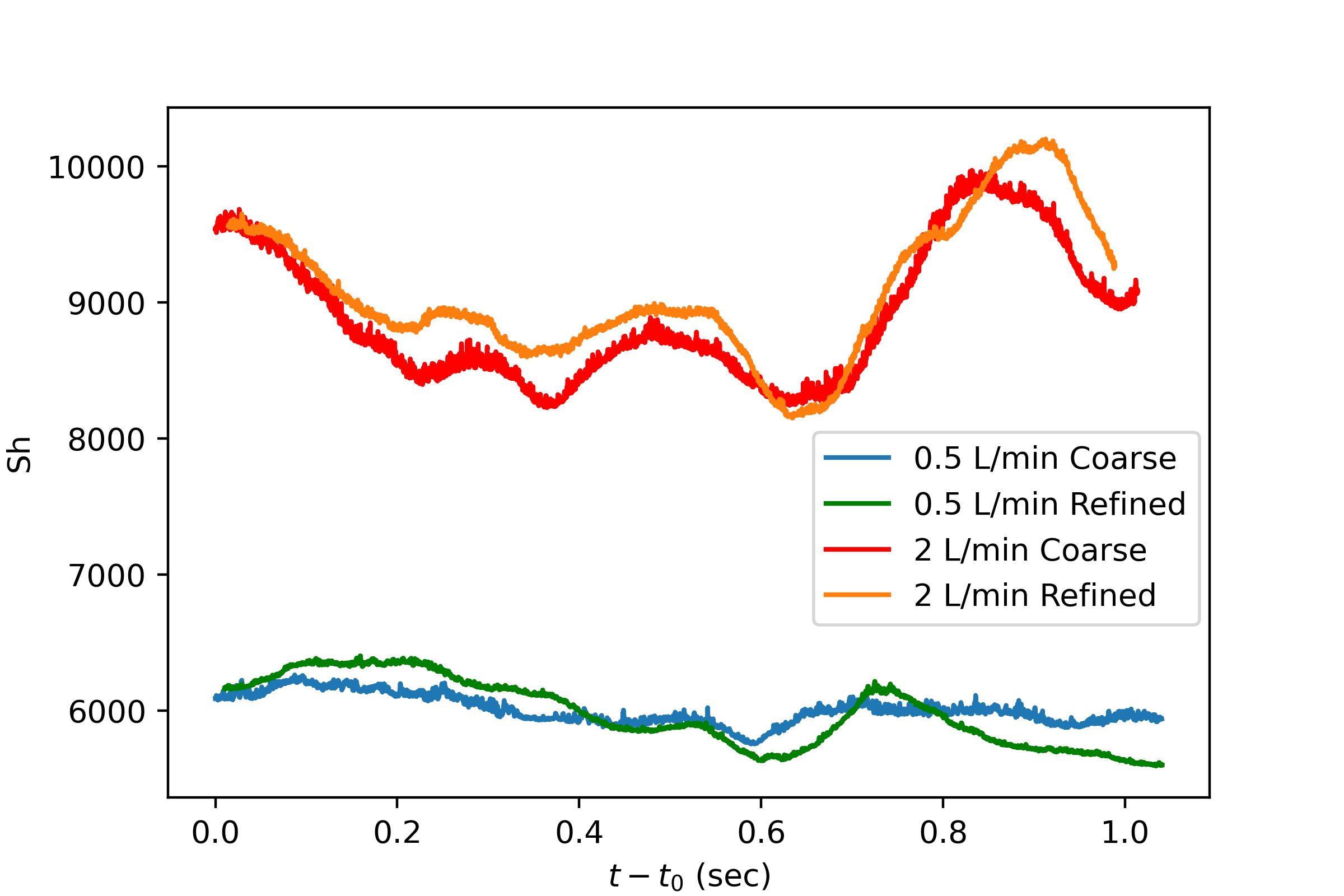}
    \caption{Effect of mesh refinement on the evolution of the Sherwood number for two flow rates of 0.5 and 2 l/min for the Kim \& Fruehan case.}
    \label{fig:mesh_conv_Sh}
\end{figure}

We compare the evolution in time of the Sherwood number from the same initial conditions for the two meshes over one second. The results are plotted in Fig.  \ref{fig:mesh_conv_Sh}, from which we observed the same trends in the instantaneous Sherwood number and less than 5 percent difference in the time-averaged Sherwood number. The coarser mesh was thus deemed to be suitable for our study and was employed in all subsequent simulations.

\subsection{Hydrodynamics results}
\begin{figure}[h]
    \centering
    \includegraphics[width=0.77\linewidth]{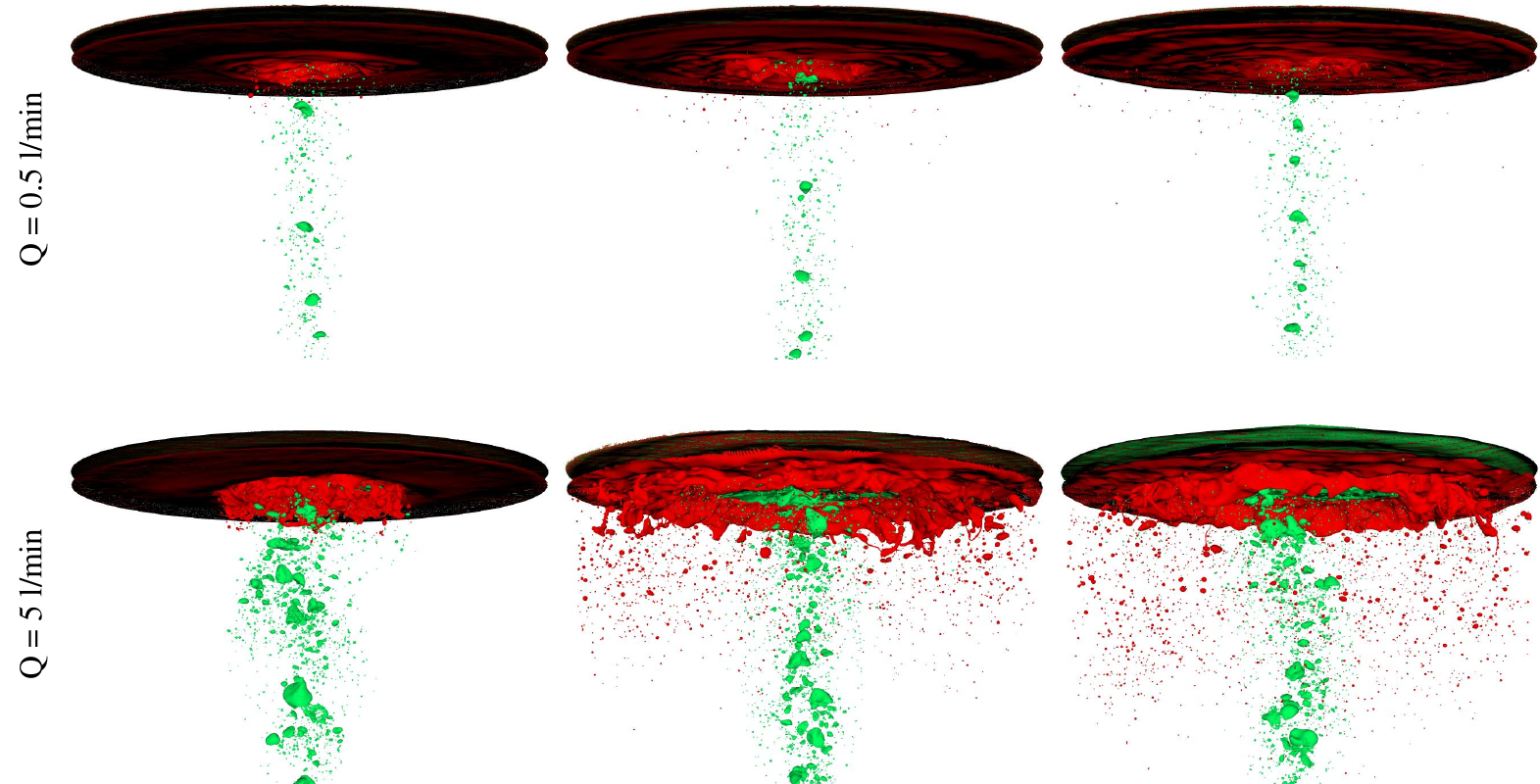}
    \caption{Side view of 3 snapshot for the lowest and highest flow rates to show the hydrodynamics of the Kim \& Fruehan case. Top row: $Q=0.5$ l/min after 3, 7.5 and 12 seconds of physical time. Bottom row: $Q=5$ l/min after 1.5, 6 and 8.5 seconds of physical time. 
    }
    \label{fig:lad_inst_hyd_side}
\end{figure}

Fig.~\ref{fig:lad_inst_hyd_side} shows the conical ladle of the Kim \& Fruehan case from a side view for the highest and lowest air injection rates investigated, namely 0.5 and 5.0 l/min, at three subsequent time steps, to illustrate the evolution of the system in time up to the reaching of quasi-steady regime. The air interface is colored in green and the oil interface in red. 
The first snapshot of each simulation shows the transient state in which the open-eye is still expanding.  
In the lower flow rate, we observe that the air bubbles rise in a cylindrical bubble column, with little mutual interaction between them, whereas at larger flow rates bubble-induced turbulence leads to a conical bubble plume. Moreover, at 5 litres per minute the oil layer atomizes and a large number of oil droplets and ligaments are found in the bulk of the water phase, an effect that was also observed in the corresponding experiment. 

\begin{figure}{H}
    \centering
    \includegraphics[width=0.77\linewidth]{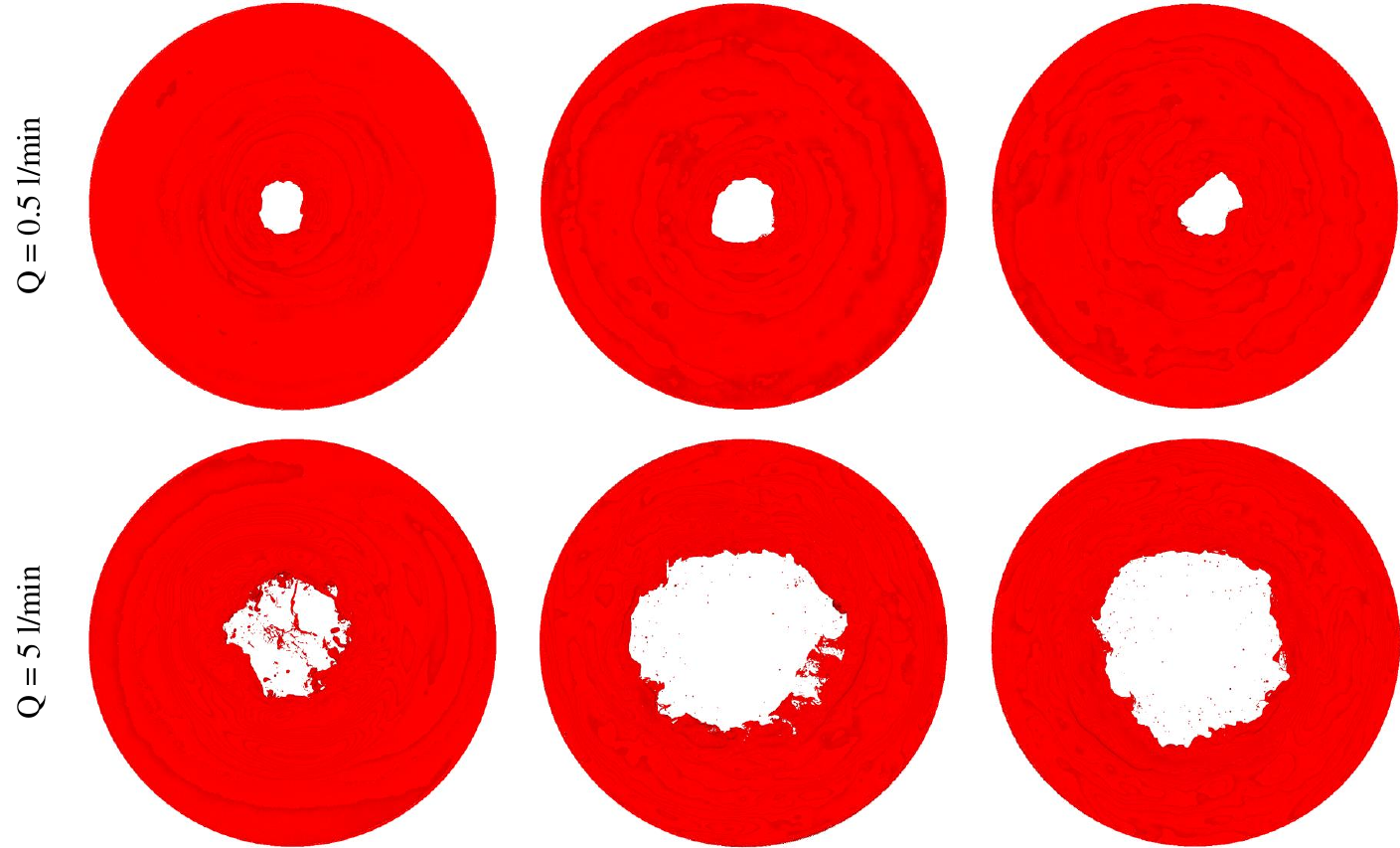}
    \caption{Top view of 3 snapshots for the lowest and highest flow rates to show the hydrodynamics of the Kim \& Fruehan case. Top row: $Q=0.5$ l/min after 3, 7.5 and 12 seconds of physical time. Bottom row: $Q=5$ l/min after 1.5, 6 and 8.5 seconds of physical time.}
    \label{fig:lad_inst_hyd_top}
\end{figure}

\begin{figure}[t]
    \centering
    \includegraphics[width=0.6\linewidth]{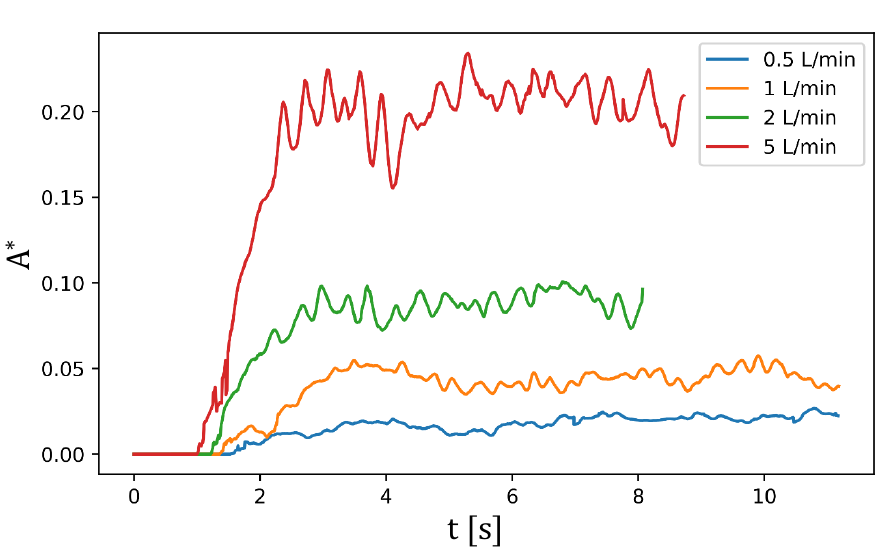}
    \caption[Evolution of the dimensionless open eye area in time]{Evolution of the dimensionless open eye area $A^*$ in time for the four flow rates in the simulations of the Kim \& Fruehan case.}
    \label{fig:openeye_evol}
\end{figure}

Fig.~\ref{fig:lad_inst_hyd_top} shows the same snapshots in a view from above, with only the oil interface portrayed in red, so that we can visualize the evolution of the open eye. To quantify its evolution, we have defined a dimensionless open eye area as
\begin{equation}
    A^*(t)=\frac{A(t)}{A_{\mathrm{ol}}^0},
\end{equation}
with $A_{\mathrm{ol}}^0$ being the initial area of the top of the oil layer. 
The evolution of $A^*$ in time is plotted in Fig.~\ref{fig:openeye_evol} for the 4 flow rates. For all flowrates, we can distinguish a transient state in which the open eye area increases steadily and then the quasi-steady regime in which its value oscillates around a plateau value with the amplitude of the oscillations depending on the flow rate. 

\subsection{Mass transfer results}
\label{sec-ladle-mass}
For what concerns the mass transfer results, the global Sherwood number is computed by directly integrating the local flux on the oil-water interface 
\begin{equation}
    \overline{Sh}(t) = -\frac{h_w}{C_{w,\infty}L_t^2 } \int_{S_{ow}(t)}  \nabla C_w (\mathbf{x},t) \cdot \mathbf{n}_\Sigma (\mathbf{x},t)  \, {\rm d}S.
\end{equation}

In the experimental campaigns, the flux is computed indirectly by measuring the concentration in the water phase over time and performing a linear regression using a lumped capacitance model \cite{joubert2021}, given by the equation 
\begin{equation}
    \frac{K_{\mathrm{w}}A}{V_{\mathrm{w}}} t= \frac{ln \left[ \frac{C_{\mathrm{w}}(t)}{C^0_{\mathrm{w}}(1+\beta) - \beta} \right]}{1+\beta},
    \label{eq:exp-kwa}
\end{equation}
where $\beta=V_{\mathrm{w}}/(PV_{\mathrm{o}})$ is a constant while $C_{\mathrm{w}}(t)$ and $C^0_{\mathrm{w}}$ are the average concentrations of the tracer in water at a general time $t$ and at $t=0$, respectively. The global Sherwood number is computed using Eq. \ref{eq:Sherwood}.

\begin{figure}[t]
    \centering
    \includegraphics[width=0.62\linewidth]{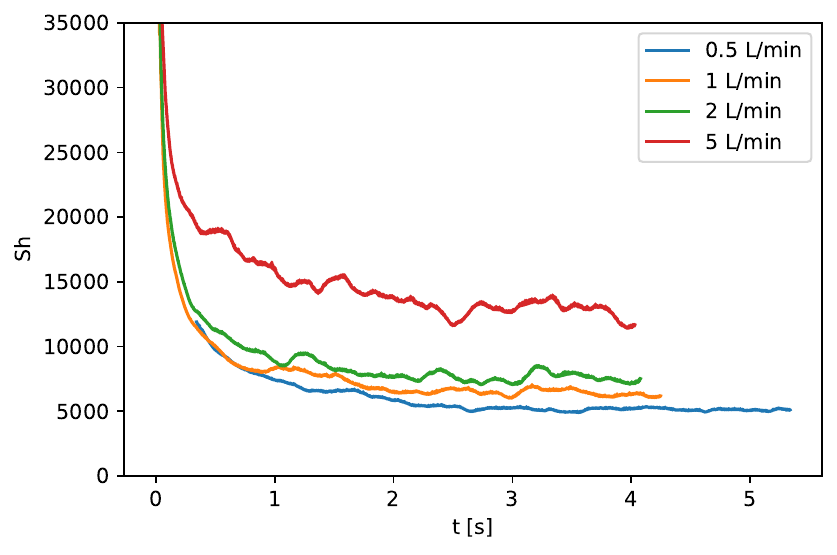}
    \caption[Evolution of the global Sherwood number $\overline{Sh}$ in time]{Evolution of the global Sherwood number $\overline{Sh}$ in time for the four mass transfer simulations of the Kim \& Fruehan case.}
    \label{fig:ladle_kwa}
\end{figure}

 The time evolution of the global Sherwood number given by Eq. \ref{eq:Sherwood}  is plotted in Fig. \ref{fig:ladle_kwa} for the four mass transfer simulations of the Kim \& Fruehan case.
The Sherwood number is initially very large and decreases over the first 2.5 seconds of simulated time, corresponding to the transient growth of a chemical boundary layer in the water phase, after which it reaches a quasi-steady state. Larger fluctuations for the global Sherwood number are observed at higher flow rates due to turbulence and atomisation of the oil layer, causing significant variations in the effective transfer area and surface renewal time \cite{doi:10.1021/ie50498a055}. The time-average and standard deviation of the Sherwood number in the quasi-steady regime are measured. 

\begin{figure}[b]
    \centering
    \includegraphics[width=0.65\linewidth]{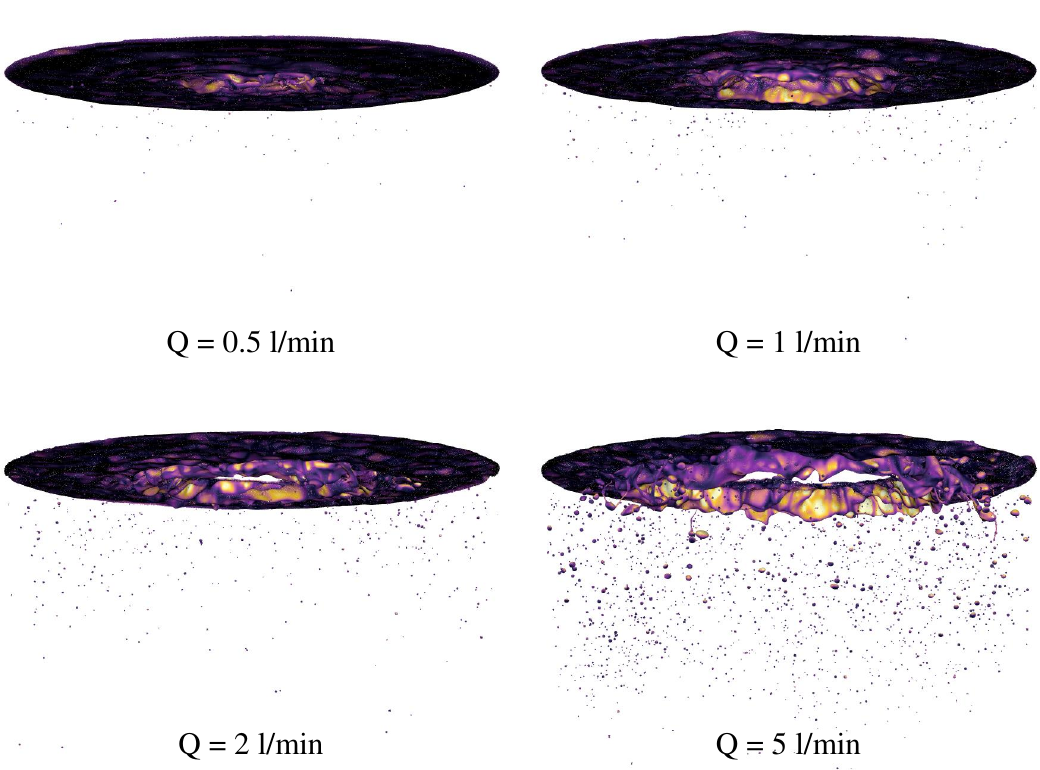} \\
    \includegraphics[width=0.5\linewidth]{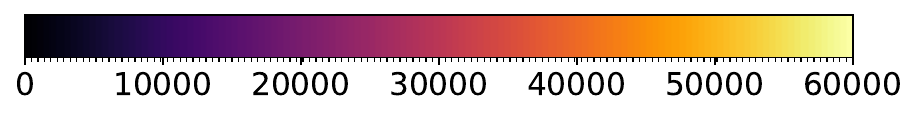}
    \caption[Sherwood number in the ladle: instantaneous side view]{Instantaneous visualisation of the Sherwood number for the four different flow rates simulated for the Kim \& Fruehan case: side view.}
    \label{fig:ladle_side_inst_sherwood}
\end{figure}

\begin{figure}[t]
    \centering
    \includegraphics[width=0.52\linewidth]{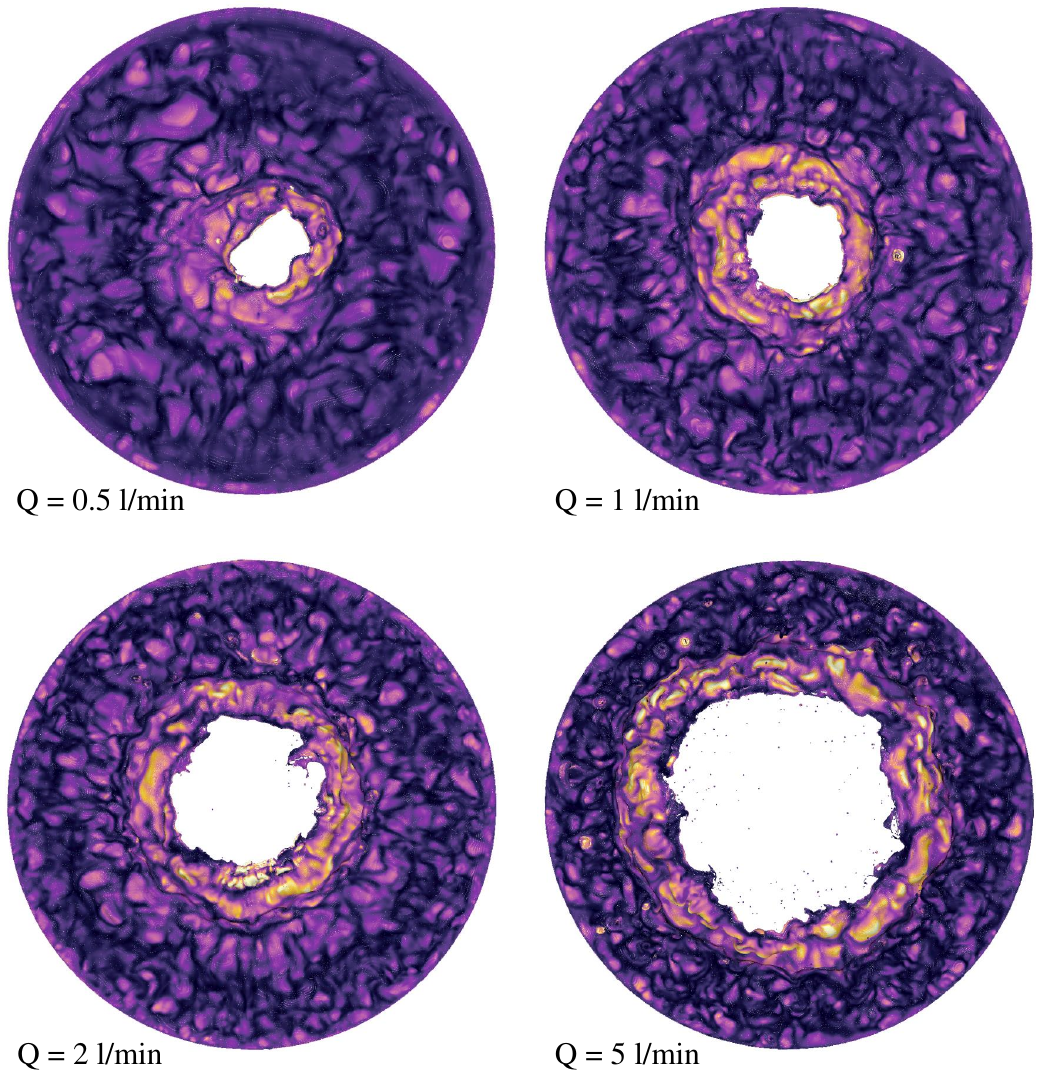} \\
    \includegraphics[width=0.5\linewidth]{figures/colorbar.pdf}
    \caption[Sherwood number in the ladle: instantaneous top view]{Instantaneous visualisation of the Sherwood number for the four different flow rates simulated for the Kim \& Fruehan case: top view.}
    \label{fig:ladle_top_inst_sherwood}
\end{figure}

In Fig.~\ref{fig:ladle_side_inst_sherwood} and Fig.~\ref{fig:ladle_top_inst_sherwood} we present the instantaneous local Sherwood number for the four different flow rates for the Kim \& Fruehan case. These figures show the oil-water interface at instants for which a quasi-equilibrium in the mass transfer has been reached in two views: one from the side and one from the top. 
We observe the highest values of the local Sherwood number
\begin{equation}
    Sh(x,t) = -\frac{h_w}{C_{w,\infty}}  \nabla C_w (\mathbf{x},t) \cdot \mathbf{n}_\Sigma (\mathbf{x},t) 
\end{equation}
in an annulus surrounding the open eye, with the maximum value of approximately 60,000. The mass flux observed in the annulus corresponds to transfer in a shear layer in the water phase. The concentration boundary layer separates at the end of the annulus and low mass transfer rates are observed in the horizontal region of the oil-water interface. The transfer rate in this region is expected to be controlled by low-amplitude interfacial turbulence. 
Interestingly, in the high flow rate case, it is observed that the very small oil drops have a relatively small effect on the global Sherwood number, which could be due to a combination of saturation and that the density ratio between oil and water is small such that buoyancy would have little effect on convection. It is measured that the total contribution to mass transfer of the detached droplets was approximately 9\% of the total mass transfer rate for a flow rate of 5 L/min. This was computed using the connected-components labelling algorithm described in \cite{basilisk_tag, HENDRICKSON2020104373}, used to segregate topologically distinct regions in a binary field. The bulk transfer, corresponding to the oil region with the largest volume,  was thus discriminated from the detached droplets (all other regions). It is shown in \ref{Appendix_Area} that the evolution of total oil-water interface area with flow rate does not correlate well with the observed mass transfer rates for the chosen flow rates. In a follow-up work, it could be interesting to investigate how the contribution to mass transfer of small droplets changes at higher flow rates.  

It also important to point out that a Sherwood number of 50,000 corresponds to a concentration boundary layer that is approximately 64 times smaller than the used grid size. This highlights the effectiveness of the SGS mass transfer model.

\begin{figure}[H]
    \centering
    \includegraphics[width=0.52\linewidth, trim={0 1.5cm 0 0},clip]{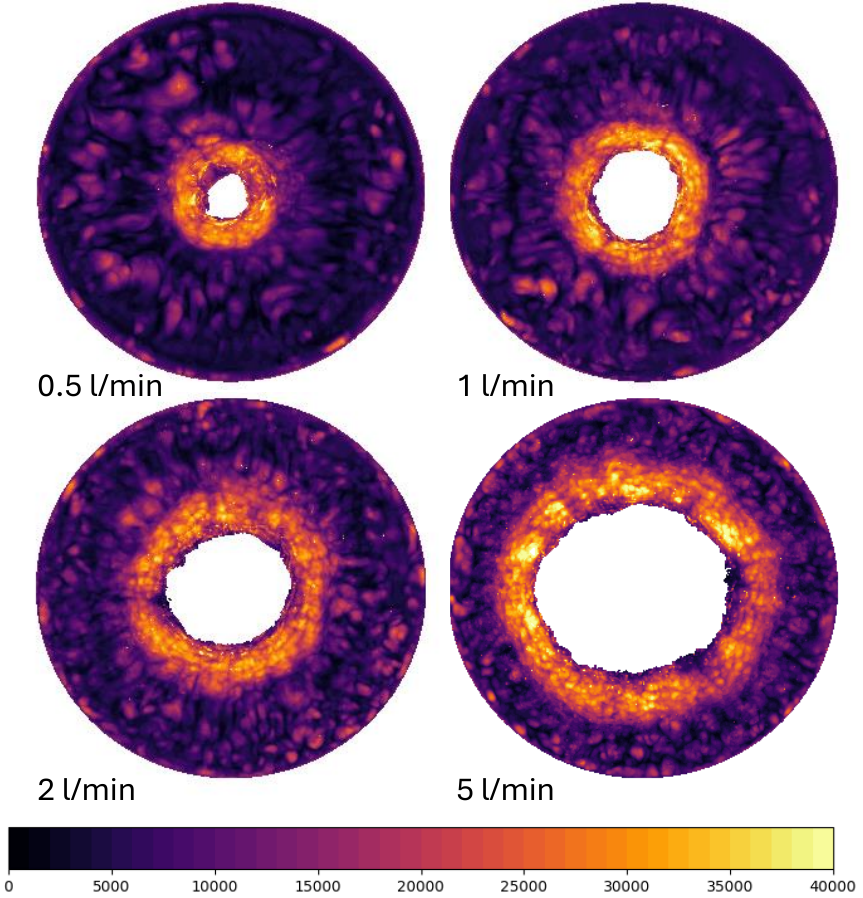} \\
    \includegraphics[width=0.5\linewidth]{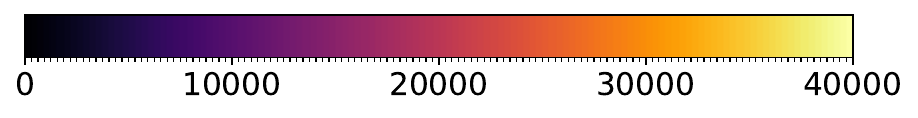}
    \caption{Time-averaged local Sherwood number $\overline{Sh}_{xz}$ for the four different flow rates simulated for the Kim \& Fruehan case.}
    \label{fig:ladle_avg_sherwood}
\end{figure}

To quantify the spatial distribution of species transfer between the bulk oil and water layers, we estimate a time-averaged local Sherwood number for each of the four simulations over a series of 20 snapshots, corresponding to 2 seconds of physical time. 
In Fig.~\ref{fig:ladle_avg_sherwood} the projection on the $xz$ plane of this average Sherwood number $\overline{Sh}_{xz}$ is shown, with the colors indicating the average value of $\overline{Sh}_{xz}$ and the white zone depicting the average position of the open eye. 
It is observed that the value of the average Sherwood number in the annulus is weakly dependent on the flow rate and that the increase in the mass transfer rate as a function of flow rate is more of a consequence of the increase in effective transfer area due to the expanded open-eye.

\section{Discussion}

\subsection{Comparison of the results between simulations and experiments}
\label{sec-ladle-comparison}

\begin{figure}[b]
    \centering
    \includegraphics[width=0.65\linewidth]{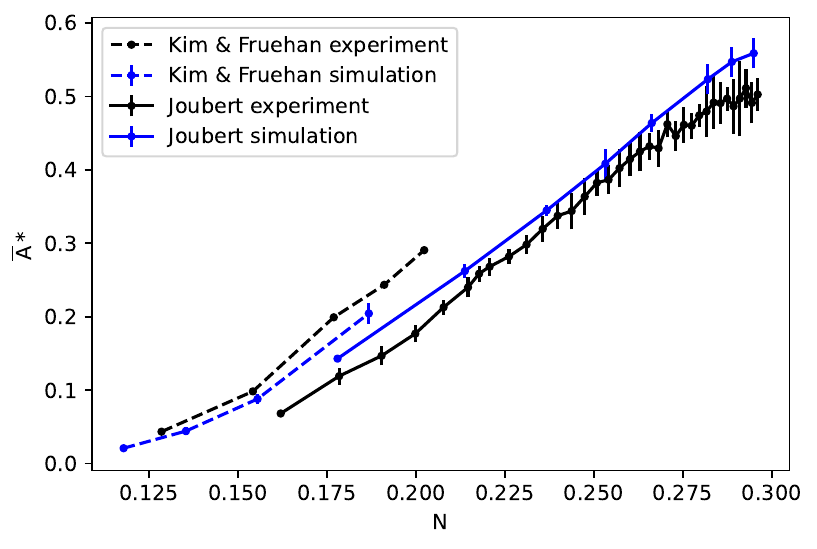}
    \caption[The dimensionless time-averaged open-eye area against modified Froude Number the simulations and experiments]{Dimensionless time-averaged open eye area $\overline{\mathrm{A}}$$^*$ against modified Froude number $N$ in the simulations and experiments. The results for the Joubert case are computed with $\Delta x \approx 264\mu$m and a post-processed minimum film thickness 1.32 mm, chosen to remove the thin capillary film. Choosing a smaller film thickness would shift the curve downwards. Error bars are not reported for the open-eye area in the Kim \& Fruehan experiment.}
    \label{fig:ladle_openeye_comparison}
\end{figure}

\begin{figure}[t]
    \centering
    \includegraphics[]{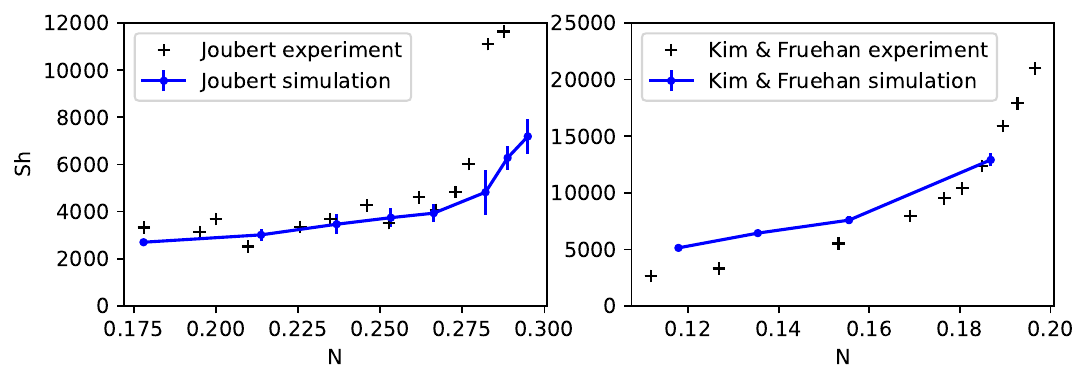}
    \caption[Simulated time-averaged global Sherwood number against modified Froude number compared with the two experiments]{Simulated time-averaged global Sherwood number ${\overline{Sh}}$ against modified Froude number $N$ compared with the experiments in \cite{joubert2021,kim1987physical}. The standard errors are used as error bars.}
    \label{fig:ladle_kwa_comparison}
\end{figure}

In this section, we compare simulation results with analogous results measured by Joubert and by Kim \& Fruehan in their experimental campaigns. This comparison is done through two macroscopic variables at pseudo-equilibrium: the time-averaged Sherwood number and the open eye area. The results are plotted in Figs. \ref{fig:ladle_openeye_comparison} and \ref{fig:ladle_kwa_comparison}.

We average the simulated open-eye area seen in Fig. \ref{fig:openeye_evol} in the steady state regime. For the simulation results of Joubert, first reported in \cite{maarekthese}, we plot them with the thin-film meniscus removed in post-processing, as the effective grid resolution without the meniscus is comparable to that used in the Kim \& Fruehan simulations. The experimental open eye area was not reported directly in the Kim \& Fruehan article but is calculated from the reported planar interfacial area.

The time-averaged Sherwood number is computed by averaging the results for the global Sherwood number reported in Fig. \ref{fig:ladle_kwa} for each flow rate in the quasi-steady regime ($ t > 2.5$ s). The global Sherwood number is calculated for the Kim \& Fruehan case using Eq. \ref{eq:Sherwood} and the reported values for $K_wA$.

The open-eye results portrayed in Fig. \ref{fig:ladle_openeye_comparison} demonstrate good agreement with both the Joubert and Kim \& Fruehan experiments. As discussed previously, we confirmed that the simulations predicts the onset of atomisation of the oil layer. It is thus demonstrated that the simulation techniques used in this work accurately capture the global hydrodynamic behaviour in the metallurgical ladle. We must qualify our results for the open-eye area with the known problem of the capillary thin oil film spreading on water that reduces the open-eye size and constitutes a source of error on likely both the experimental and simulated open-eye measurements. This film is difficult to capture in the experiments, since it is not clearly distinguishable in the acquired pictures, while in the simulations it leads to mesh dependent results. In the current simulations, we remove the meniscus from our measured simulated open-eye with post-processing, but in the future it would be better to set a minimum thin-film thickness using the Manifold Death algorithm \cite{CHIRCO2022111468}, which perforates thin sheets in an interface described by a VOF field at a user-defined thickness. The film-thickness parameter would be chosen from an estimated measurable film-thickness in experiments. We also have to point out that, according to the surface tension values for the argon-slag-steel system reported in \cite{jothese, lachmund2003slag}, slag is not expected to spread on liquid steel, thus removing this source of error when simulating the industrial system, but also showing a difference between the industrial and water experiments.

From Fig.~\ref{fig:ladle_kwa_comparison}, we observe the same qualitative behaviour between the experiments and simulations with approximately the same slope of the dependence of $\overline{Sh}$ on $N$ in the first regime of mass transfer. It is observed in the Joubert case that the simulations predict a change in mass transfer regime near the experimental critical Froude number, though the mass flux in the second regime is under-predicted compared to the experiments. 
Quantitative agreement is observed between the simulations and experiments in the first regime of mass transfer for the Joubert case, whereas the simulations over-predict the mass transfer rate at low flow rates compared to experiments for the Kim \& Fruehan case. 
The over-prediction can be attributed to factors in both the simulations and the experiment. In the case of the former, the use of the error function as the boundary layer profile leads to a maximum in the predicted flux compared to the possible profiles from Eq. \ref{ediff}, with the other extreme L{\'e}v{\^e}que profile expected to decrease the predicted flux by approximately 12 percent \cite{maarekthese}. 
It is also possible that the relatively coarse grid under-predicted the viscous dissipation in the oil phase. Lastly the difference in measured time between experiments (hours) versus simulations (seconds) means it is possible we are missing effects that occur at large time-scales e.g. the saturation of the oil layer.
In the case of the experiments the inconsistent fluid volumes and missing measured surface tension and oil diffusivity lead to a likely small source of error. The most plausible explanation for the difference between the simulation and experiments is the position of the concentration probe used in the experiment, described in \cite{kim1987physical} to be 1 cm away from the bottom and near the wall of a model ladle. It seems likely that the tracer is not well-mixed in this region, with the concentration evolution of the probe lagging behind the average concentration evolution in the ladle, thus introducing an error when using the lumped capacitance model. In future experimental/simulation campaigns, the concentration  in the water phase should be measured in experiments with passive measurement techniques (e.g. laser interferometry) in several places in the water such that an uncertainty on the experimental mass flux can be included. It is also possible, though potentially less preferable, to include simulated concentration probes directly in the CFD simulations.

\subsection{Scalability of the model}

As a last remark, we discuss the scalability of the described numerical methods from simulating a reduced water experiment to the industrial steel-slag-argon system. From a modeling perspective, the principal difference between the used mathematical model and the physical system is that it is known that the compressibility of air can no longer be ignored \cite{jothese, maarekthese}. This could affect the dynamics in the bubble plume, though its effect on macroscopic quantities such as the open-eye size and species mass flux is uncertain. More importantly, the Reynolds number of the industrial system increases from $10^4$ to $10^6$ when compared to the reduced system. Using the scaling argument of Pope \cite{Pope_2000}, which says that the computational cost of a turbulent system scales with $Re^{11/4}$, estimates that simulating the industrial system would cost roughly 300 thousand times more than the reduced system. Using  adaptive mesh resolution (AMR) generally adjusts the coefficient rather than the exponent and it can be assumed the AMR cell reduction in the model system would be of the order of that seen in the industrial system. As the reduced scale simulations are already expensive, ($\sim 10^5 \mathrm{\; to } \; 10^6$ CPU-hours), simulating the industrial system is beyond the computational feasibility required for the current approach to be integrated into the engineering design cycle. To combat this, one could introduce LES-based turbulence modelling, which has been successfully applied to simulate similar Reynolds number flows at moderate costs \cite{goc2021large}. Furthermore the morphological complexity of the system could be simplified by using morphological-multiscale approaches. The simplest example could be to replace the interface-resolved bubble plume with an Euler-Euler model, using a tracer to represent the local buoyancy and a bubble turbulent Schmidt number to simulate the dispersion of bubbles. A more challenging improvement would be to use a combination of interface-resolved methods and Euler-Lagrange approaches to resolve the large interfaces and use Lagrangian particles to model the small detached droplets of the slag layer \cite{KASBAOUI2025105175}. The mass flux from the Lagrangian particles would be computed using empirical correlations. Lastly switching to more advanced high performance computing technologies (e.g. GPU-accelerated computing) could save roughly one order of magnitude on energy consumption \cite{gustafsson2025cfd}. Importantly, all of the described improvements are existing technologies which are compatible with the methods used in this paper.  

\section{Conclusions}
In this paper we have investigated hydrodynamics and mass transfer in two reduced-scale model metallurgical ladles by means of direct numerical simulation (DNS), comparing our results with experimental data from the literature. 
The numerical method presented shows a significant advancement on the state of the art for computational mass transfer analysis in reduced-scale metallurgical ladles. 
The results shown demonstrate that combining multiscale approaches with adaptive mesh resolution to treat ranges of length scales in the hydrodynamic system together with a subgrid-scale (SGS) model to treat the thin concentration boundary layers, yields relative agreement between experiments and simulations. To the best of the authors knowledge, this is achieved for the first time without extrapolating results simulated with artificially high diffusivities, as the SGS model allowed us to simulate very large $Sc$ numbers with computational costs comparable to simulating just hydrodynamics without mass transfer.

Having good agreement between simulations and experiments is crucial, as the former can give much richer data than what is measurable in experiments. For example, we can directly measure the time-averaged velocity field to understand physical phenomena present in the system rather than rely on macroscopic, optically-measurable quantities such as the open-eye size. We can also directly measure the local distribution mass flux at the oil-water interface, which gives a better understanding of the relationship between the input flow rate and the output flux. We can use this rich data to inform a phenomenological model that predicts the measured quantities as a function of the flow rate. A mechanistic model that explains the measured quantities and attempts to unify all hydrodynamic and mass transfer experiments is under development and will be the subject of a future paper.  

The described work and the possible avenues of improvement represent a big step in the numerical mass transfer simulation of industrial systems such as the secondary refining ladle, which should be used as part of the metallurgical engineering design cycle going forward.

\vspace{\baselineskip}

\appendix

\section{Structure of the bubble plume considering bouyant and momentum driven effects}

\label{bubble_plume_structure}

We consider the relative effects  of bouyancy and momentum-driven effects on the structure of the bubble plume. We assume the classical scaling for the Boussinesq buoyant plume with the rise velocity given by \cite{monin1971statistical,Tennekes:72}
    \begin{equation}
        {\overline  u_z} (z)  \sim \left(\frac{g Q} z \right)^{1/3}.
    \end{equation}
For a self-similar free jet, the velocity follows \cite{monin1971statistical,Tennekes:72}
\begin{equation}
      {\overline  u_z} (z)  \sim \frac { ( M / \rho_w)^{1/2}}z \label{muzscale} 
\end{equation}
 where the momentum flux $M$ is given by
\begin{equation}
    M = \int_{S_z}  \langle  \rho u^2_z \rangle  {\mathrm{d}}S 
\end{equation}
 Assuming a plug flow in the inlet, the momentum flux may be estimated at the inlet by
 $M= \rho_a  \langle V_I^2 \rangle \pi d_I^2/4$ where $V_I$ is the inlet velocity,
 related to the flow rate by $Q/(4 \, d_i^2)$, leading to relation between momentum flux and volume flux
 \begin{equation}
      M = \frac{4 \rho_a}{\pi d_I^2} Q^2
 \end{equation}
 and the velocity scaling 
\begin{equation}
     {\overline  u_z} (z)  \sim   \left(\frac{4 \rho_a}{\pi \rho_w}\right)^{1/2} \frac Q {d_I z} 
\end{equation}
A rough approximation may obtained by combining the two scalings 
\begin{equation}
      {\overline  u_z} (z)  \simeq c_{u,z} \left(\frac{g Q} z \right)^{1/3} + c_{u,z,m}  \left(\frac{4 \rho_a}{\pi \rho_w}\right)^{1/2} \frac Q {d_I z}, \label{uzscale} 
\end{equation}
where we have introduced constants $c_{u,z}$ and $c_{u,z,m}$. The ratio of the plume (buoyancy-driven) velocity and of the free-jet (momentum-driven)
velocity, that is the ratio of the two terms of Eq.
(\ref{uzscale}) at height of the top of the water layer $h_w$ is
\begin{equation}
    R =  \frac {h_w} {d_I} \left( \frac{Q^2}{g h_w^5} \right)^{1/3}  \left( \frac{4 \rho_a} {\pi \rho_w}\right)^{1/2} \label{UfUp}
\end{equation}

For all flow rates chosen in the two test-cases, the value computed by Eq. \ref{UfUp} is small, showing momentum-driven effects are negiligble on the development of the structure of the bubble plume.

\section{Relationship between the oil-water interface area and mass transfer} \label{Appendix_Area}

\begin{figure}[t]
 \centering
    \begin{minipage}[c]{.49\textwidth}
      \centering
      \includegraphics[width = \textwidth]{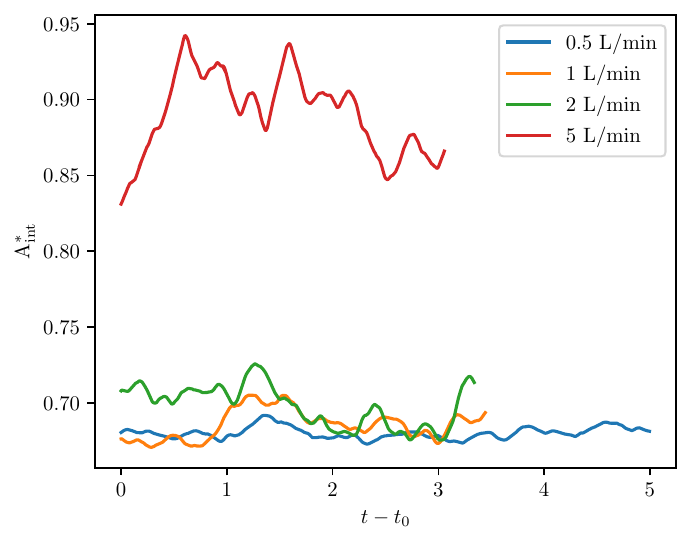}
     \subcaption{}
    \end{minipage}
    \begin{minipage}[c]{0.49\textwidth}
      \centering
      \includegraphics[width = \textwidth]{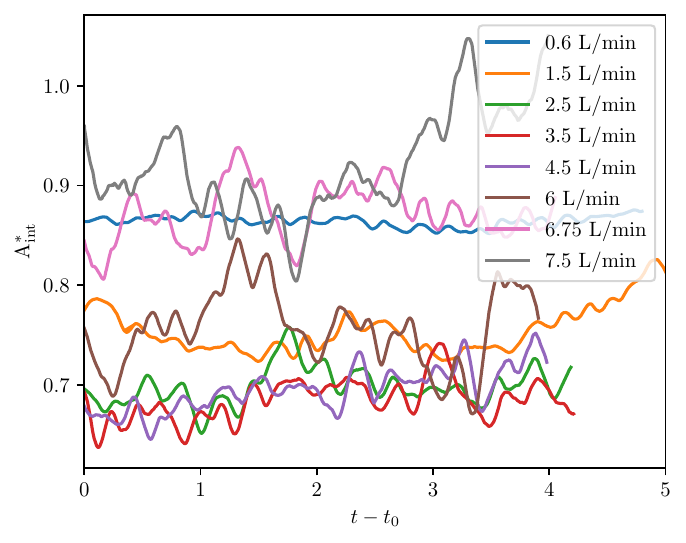}
      \subcaption{}
    \end{minipage}
    \caption{Instantaneous nondimensional oil-water interfacial area computed in the quasi-steady regime for (a) the Kim \& Fruehan simulations and (b) the Joubert simulations}
    \label{inst_A_int}
\end{figure}

\begin{figure}[t]
    \centering
    \includegraphics[width=0.6\linewidth]{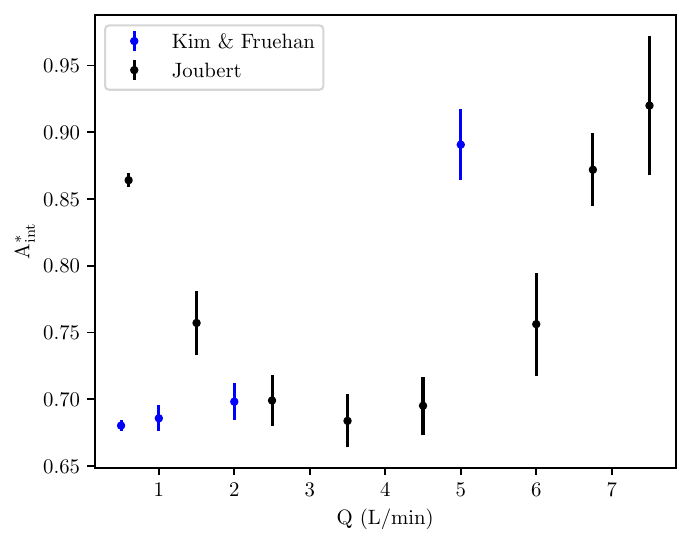}
    \caption{Time-averaged nondimensional oil-water interfacial area for both sets of simulations. The standard error is used as error bars.}
    \label{avg_A_int}
\end{figure}

We consider the relationship between the oil-water exchange area and mass transfer. This follows the principle of the interfacial area concentration, a geometric quantity commonly used in mixture models of bubbly flows to characterize mass, momentum and energy transfer, usually characterized as the interfacial area per unit volume \cite{KATAOKA1990163}. 
We assume that the height of the water layer has little effect on the total oil-water interfacial area, so that instead we define an order one non-dimensional quantity $\mathrm{A_{int}^*}$, given by the total oil-water interfacial area divided by the cross sectional area of the tank evaluated at the height of the oil layer. We plot both instantaneous and time-averaged values of the non-dimensional interfacial area in Fig. \ref{inst_A_int} and \ref{avg_A_int}. We observe that in the Kim \& Fruehan case, the interfacial area only increases slightly at the lowest flow rates and then increases dramatically for the highest flow rate, corresponding to the atomising regime. More strikingly in the Joubert case, the open-area decreases with increasing flow rates in the non-atomising regime before increasing when the oil layer atomises. To explain this phenomenon, we consider a simplified approximation of the oil-water surface area without considering fragmentation
\begin{equation}
    \mathrm{A_{int}} = \mathrm{A_{cr}} - \pi r_{\mathrm{oe}}^2 + 2\pi  r_{\mathrm{oe}} h,
\end{equation}
where $\mathrm{A_{cr}}$ is the cross sectional area, $r_{\mathrm{oe}}$ is the open-eye radius and $h$ is the height of the oil layer given by 
\begin{equation}
    h = \frac{V_o}{\mathrm{A_{cr}} - \pi r_{\mathrm{oe}}^2},
\end{equation}
where $V_o$ is the oil volume.
As the open-eye radius increases, the total exchange area would thus decrease. This phenomenon is not quite observed in the case of the Kim \& Fruehan case, which is caused both by the open-eye area being larger relative to the tank cross-sectional area, consequently causing a less significant effect, and by the larger interface deformation and smaller amounts of oil fragmentation observed at intermediate flow rates for these simulations.  

Since the evolution of the interfacial area in the non-atomising regime does not correspond to the observed increase in mass transfer rate, it is demonstrated that the interfacial area does not correlate very well with the mass transfer rate in the ladle experiments, at least for low to intermediate flow rates. 
This is different to what is commonly observed for mass transfer in turbulent bubbly flows, where it can often be assumed that the Sherwood number of each bubble is either statistically uniform or given by a known distribution. Consequently, the global Sherwood number would be the product of the mean bubble Sherwood number and the interfacial area.

In the ladle case, we can consider three topologically distinct regions: the annulus around the open-eye, the relatively flat region between the oil and water bulks outside the annulus, and all the detached droplets. Each region is governed by different flow and mass transfer phenomena, namely forced convection in a shear layer for the annulus, free convection for the outer region, and droplet mediated mass transfer for the droplets. It is observed that the Sherwood number in the annulus region has a dominant role in the overall mass transfer rate, at least for the flow rates considered. It is expected that at higher flow rates than the ones simulated, mass transfer by fragmented droplets would play a more significant role, thus corresponding to the shift in mass transfer regimes observed at high flow rates in experiments.

\clearpage

\acknowtext
We thank the European PRACE group, the Swiss supercomputing agency CSCS, the French national GENCI supercomputing agency and the relevant supercomputer centres for their grants of CPU time on massively parallel machines, and their teams for assistance and the use of Irene-Rome at TGCC and ADASTRA at CINES.
This project was provided in particular with HPC computing and storage resources by GENCI at TGCC thanks to the grants 2024 - A0172B14629, 2023-A0152B14629 and 2022 SS012B15405 on the supercomputer Joliot Curie’s SKL and Irene-Rome partitions.

\funding
This research project was funded by the European Research Council (ERC) through the European Union’s Horizon 2020 Research and Innovation Programme (Grant Agreement No. 883849 TRUFLOW).

\conflict
The authors have nothing to disclose.

\dataavailability

Data will be made available on request.

\authorcontrib

S.D.R: conceptualization, formal analysis, investigation, methodology, software, validation, visualization, writing—original draft, writing—review and editing; J.M: conceptualization, formal analysis, investigation, methodology, software, validation, visualization, writing—original draft, writing—review and editing; S.Z.: conceptualization, formal analysis, funding acquisition, investigation, methodology, project administration, resources, software, supervision, validation, writing—review and editing. All authors gave final approval for publication and agreed to be held accountable for the work performed
therein.









\supplementary

In the supplementary material the reader can find additional plots and figures for the Joubert case. We also include videos of the performed simulations where the oil and air interfaces are visualized.


\bibliographystyle{bibstyle.bst}  
\bibliography{bibliography}  



%
%
%
%
%
%


\end{document}


\textbf{SUPPLEMENTARY MATERIALS}

\section{Source code}

The Basilisk source code can be found here http://www.basilisk.fr/src/. Header files and the simulation setup file can be found in the Basilisk sandbox http://basilisk.fr/sandbox/jmaarek/ladle/.

\section{Supplementary figures for the Joubert case}

In the main paper, for the sake of brevity some plots were only presented for the Kim \& Fruehan case. In this section, the analogous plots for the Joubert case are provided.

\begin{figure}[h!]
    \centering
    \includegraphics[width=0.8\linewidth]{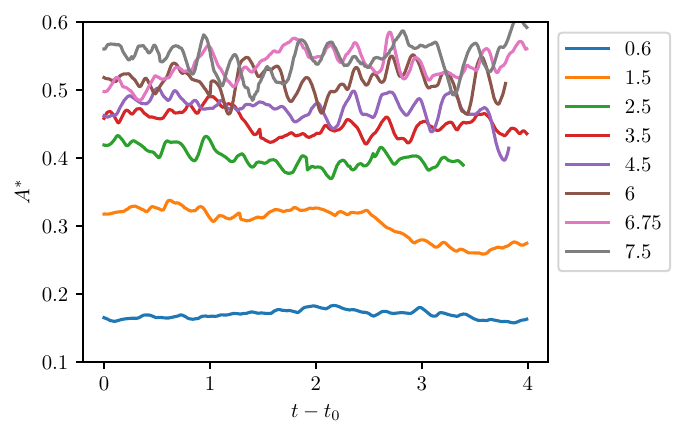}
    \caption{
    Evolution of the simulated dimensionless open eye area $A^*$ for 4 seconds of during the quasi-steady regime. Results are given for all simulated flow rates in L/min.} 
    \label{fig:openeye_evol_cube}
\end{figure}

Figure \ref{fig:openeye_evol_cube} shows the time evolution of normalized open-eye area for the simulated flow rates in the quasi-steady regime. Results given are for a grid size $\Delta =$ 512 $\mu$m without removing thin films in post-processing.

\begin{figure} [h!]
    \centering
    \includegraphics[width=0.65\linewidth]{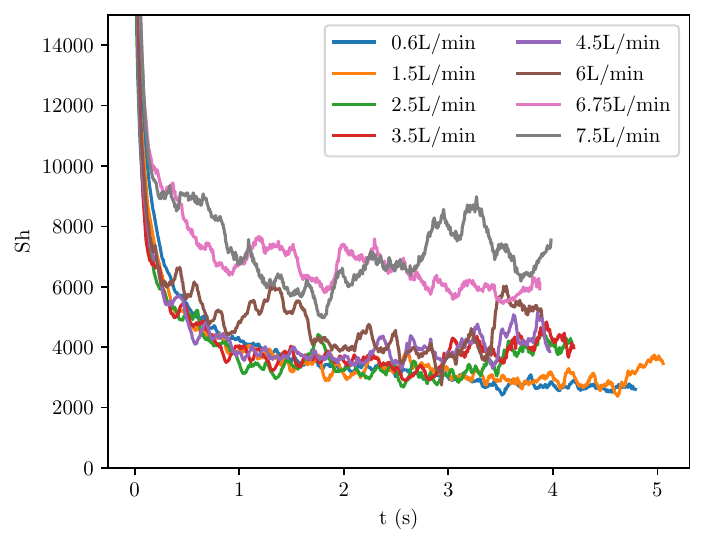}
    \caption{Evolution of the global Sherwood number $\overline{Sh}$ in time for the four mass transfer simulations of the Joubert case.
    }
    \label{fig:ladle_kwa_cube}
\end{figure}

The time evolution of the global Sherwood number is plotted in Fig. \ref{fig:ladle_kwa_cube} for simulated flow rates. It is observed that the 6.75 and 7.5 L/min are significantly above the rest, where the atomizing regime is present.

\begin{figure} [h]
    \centering
    \includegraphics[width=0.6\linewidth]{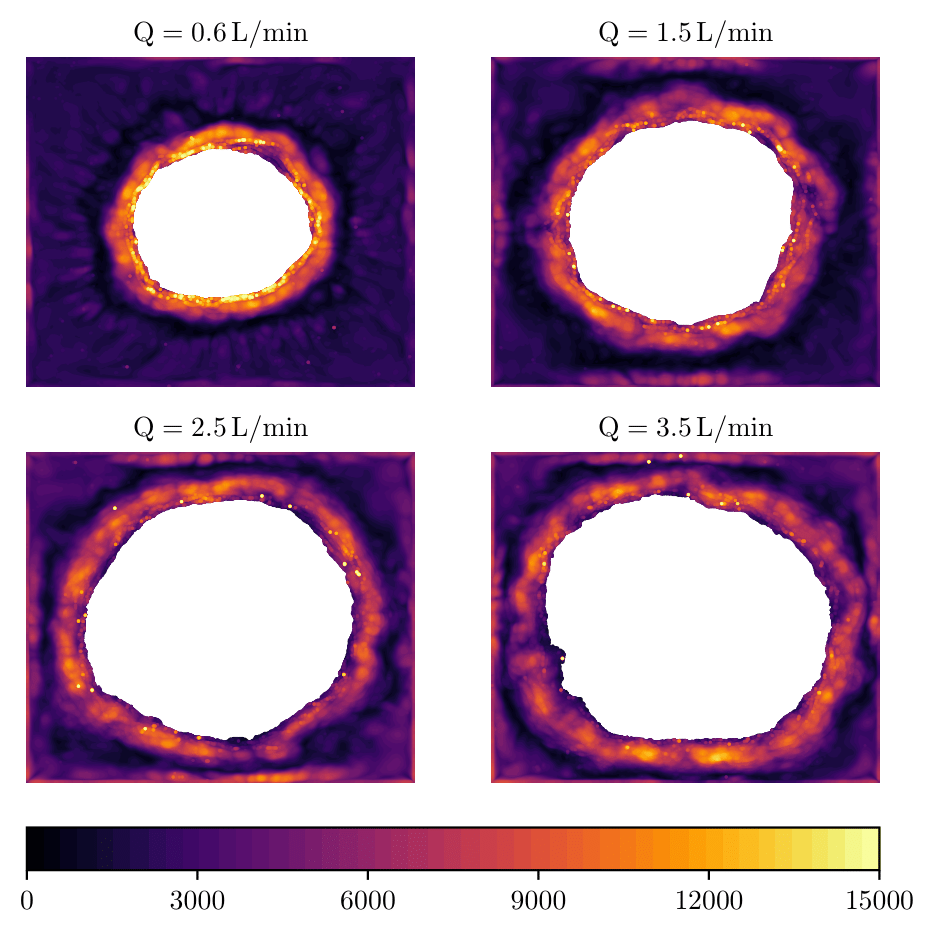}
    \caption{Time-averaged local Sherwood number $\overline{Sh}_{xz}$ for the four different flow rates simulated for the Joubert case.}
    \label{fig:ladle_avg_sherwood_cube}
\end{figure}

The time-averaged local Sherwood number for four simulated flow rates below the atomizing critical flow rate are shown in Fig.~\ref{fig:ladle_avg_sherwood_cube}. The time-average is computed from a series of 20 snapshots corresponding to 2 seconds of physical time. The projection on the $xz$ plane of this average Sherwood number $\overline{Sh}_{xz}$ is portrayed, in an analogous manner as what was done for the Kim \& Fruehan case in the main paper.

In the file Ladle\_Movies.zip, videos of the performed simulations are included, with the oil and air interfaces visualized. The videos were made at 100 fps. Note in the videos of the Kim \& Fruehan case the interface in the solid region can be ignored.

